\documentclass[12pt]{iopart}

\usepackage{todonotes}
\usepackage{graphicx}
\usepackage{cite}  
\usepackage{amssymb}
\usepackage{bm}
\usepackage{xcolor,xspace}
\usepackage{wasysym}  
\usepackage{hyperref}
\hypersetup{colorlinks=True,urlcolor=blue,linkcolor=blue,
citecolor=blue,filecolor=black}
\usepackage[capitalise]{cleveref}  
\usepackage{braket}
\usepackage{array}
\usepackage[percent]{overpic}  
\newcolumntype{L}{>{$}l<{$}} 
\newcommand{\eq}{\!=\!}
\newcommand{\be}[4]{B(E2;#1^+_{#2} \!\to\! #3^+_{#4}) } 

\begin{document}

\title{Intertwined quantum phase transitions in 
the zirconium and niobium isotopes}

\author{N. Gavrielov}
\address{GANIL, CEA/DSM–CNRS/IN2P3, Bd Henri Becquerel, BP 
55027, F-14076 Caen Cedex 5, France}
\eads{\mailto{noam.gavrielov@ganil.fr}}

\vspace{10pt}
\begin{indented}
\item[]February 2024
\end{indented}

\begin{abstract}
Nuclei in the $A\approx100$ region exhibit intricate 
shape-evolution and configuration crossing signatures. 
Exploring both even-even and their adjacent odd-mass nuclei 
gives further insight on the emergence of deformation and 
shape-phase transitions. We employ the algebraic frameworks 
of the interacting boson model with configuration mixing 
and the new interacting boson-fermion model with 
configuration mixing in order to investigate the even-even 
zirconium with neutron number 52--70 ($^{40}$Zr) and 
odd-mass niobium ($_{41}$Nb) isotopes with 52--62. We 
compare between the evolution in energy levels, 
configuration and symmetry content of the wave functions, 
two neutron separation energies and $E2$ transition rates. 
The comparisons between the two chains of isotopes denote 
the occurrence of intertwined quantum phase transitions 
(IQPTs) in both chains. Such a situation occurs when two 
configurations, normal and intruder, cross through the
critical point of a Type~II quantum phase transition (QPT), 
and the intruder configuration undergoes on its own a 
Type~I shape-evolution QPT from a spherical shape (weak 
coupling scenario) to axially deformed rotor (strong 
coupling scenario) in the Zr (Nb) isotopes.
\end{abstract}


\section{Introduction}\label{sec:intro}
Nuclei in the region of $A\approx100,Z\approx40$ have 
said to undergo an abrupt change in the structure of their 
ground and non-yrast states as one varies the number of 
neutrons \cite{Heyde2011, Garrett2022}. 
The situation has been ascribed to multiple shell model 
configurations that mix through the proton-neutron 
interaction and cross \cite{Federman1979}, while also 
changing their intrinsic shape \cite{Togashi2016}. Such 
scenarios, shape evolution and configuration crossing, have 
been termed as, respectively, Type~I and type~II quantum 
phase transitions (QPTs).

QPTs are zero temperature phase transitions that occur from 
quantum fluctuations as a function of coupling constants 
in the Hamiltonian, which serve as the control parameters 
\cite{Cejnar2010}. 
In nuclear physics, where this concept has been presented 
for the first time \cite{Gilmore1978a, Gilmore1979}, an 
intrinsic shape parameter serves as the order parameter and 
thus one terms it often shape-phase transitions. The 
concept of QPTs in atomic nuclei has been separated into 
two Types, I and II, which are said to occur in different 
situations. 

\begin{figure}
\centering
\includegraphics[width=0.6\linewidth]{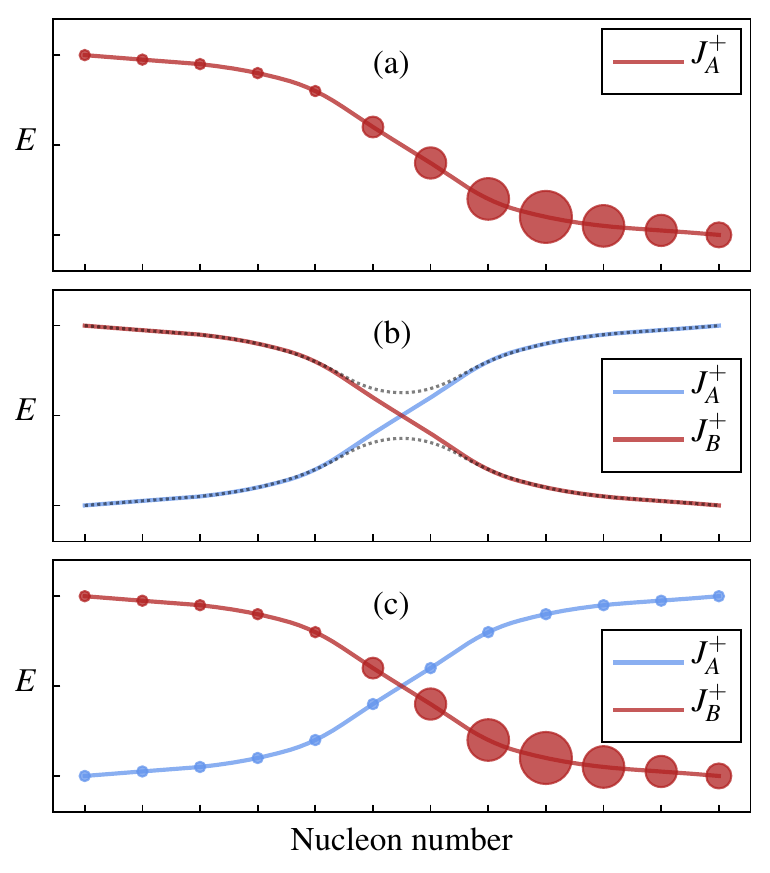}
\caption{Schematic illustration for the evolution with 
nucleon number of energies (in arbitrary units) of the 
lowest $J^{+}$ states of one or two configurations, A and 
B. (a)~Type~I QPT: shape changes within a single 
configuration (small and large circles denote weak and 
strong deformation, respectively). {(b)~Type~II QPT:} 
coexisting and possibly crossing of two configurations, 
usually, with strong mixing. The dashed lines depict the 
mixing, as in a two states mixing scenario.
(c)~IQPTs: abrupt crossing of two configurations, with weak 
mixing, accompanied by a pronounced gradual shape evolution 
within each configuration.\label{fig:iqpt}}
\end{figure}
The first QPT, Type~I, is considered when nucleons are 
added in a single shell model configuration and a shape 
evolution occurs as one adds more nucleons, as shown 
schematically in \Fref{fig:iqpt}(a). The circles denote the 
amount of deformation and as more nucleons are added it 
increases, which lowers the absolute energy of the ground 
state. A Type~I QPT can be described by the following 
Hamiltonian 
\begin{equation}\label{eq:type-i}
\hat H = (1-\xi)\hat H_1 + \xi\hat H_2~.
\end{equation}
As the control parameter $\xi$ varies from 0 to 1, the 
equilibrium shape and symmetry of the Hamiltonian vary from 
those of $\hat H_1$ to those of $\hat H_2$. 

The second QPT, Type~II, is considered when nucleons are 
added in multiple shell model configurations that interact 
and cross. This is in the sense that higher lying states 
that are associated with nucleons residing in a higher 
configuration are lowered to become the ground state. This 
is shown schematically in \Fref{fig:iqpt}(b) where one 
observes what appears to resemble a two state mixing 
scenario. In this case, Type~II QPTs can be described by a 
matrix Hamiltonian \cite{Frank2006}
\begin{equation}\label{eq:type-ii}
\hat H =
\left [
\begin{array}{cc}
\hat H^{\rm A}(\xi^{\rm(A)}) & \hat W(\omega) \\ 
\hat W(\omega)  & \hat H^{\rm B}(\xi^{\rm(B)})
\end{array}\right] 
\end{equation}
given here for two configurations, denoted by the indices 
$A$ and $B$, and $\hat W$ denotes their coupling.

Recently, studies have shown the occurrence of such 
crossing and shape evolution in the even-even zirconium 
($_{40}$Zr) isotopes \cite{Gavrielov2019, Gavrielov2020, 
Gavrielov2022} and the adjacent odd-$A$ niobium 
($_{41}$Nb) isotopes \cite{Gavrielov2022c, Gavrielov2023a}. 
This unique scenario has been termed as intertwined quantum 
phase transitions (IQPTs), which denotes the identification 
of both Types of QPTs, I and II, in the same chain of 
isotopes. 

\section{Theoretical framework}
For the study of QPTs involving configuration crossing we 
shall use the algebraic framework of the interacting boson 
model with configuration mixing \cite{IachelloArimaBook, 
Duval1981, Duval1982} for the even-even Zr isotopes 
and the interacting boson-fermion model with configuration 
mixing for the odd-mass Nb isotopes \cite{Gavrielov2022c, 
Gavrielov2023a}.
\subsection{IBM-CM}
The IBM for a single shell model configuration has been 
widely used \cite{IachelloArimaBook} to describe low-lying 
collective states in nuclei in terms of $N$ 
monopole ($s$) and quadrupole ($d$) bosons, representing 
valence nucleon pairs.
The Hamiltonian interactions are Hermitian, 
rotationally invariant and conserve the total 
number of $s$ and $d$ bosons,
\begin{equation}\label{eq:boson-number}
\hat N=\hat n_s+\hat n_d = s^\dagger s+\sum_\mu 
d^\dagger_\mu d_\mu~.
\end{equation}
The number of bosons \eref{eq:boson-number} is fixed by the 
microscopic interpretation of the IBM 
\cite{IachelloTalmi1987} to be the total number of proton 
and neutron particle or hole pairs counted from the nearest 
closed shell.

For excitations involving $n$-particles and $m$-holes 
($n$p-$m$h) from multiple shell model configurations, one 
can extend the IBM to IBM with configuration mixing 
(IBM-CM) by associating each $n$p-$m$h excitations to a 
different $N$-boson space, such as 0p-0h ($N$), 2p-2h 
($N+2$), 4p-4h ($N+4$), etc. In such a case, the 
Hamiltonian named $\hat H_b$ is written in matrix form, 
which for two configurations is written as in 
\cref{eq:type-ii}, where $\hat H_b^{\rm A}(\xi^{(\rm A)})$ 
represents the normal configuration ($N$ boson space, 
denoted henceforth by A) and $\hat H_b^{\rm B}(\xi^{(\rm 
B)})$ represents the intruder configuration ($N\!+\!2$ 
boson space, denoted henceforth by B), corresponding to 
2p-2h excitations across the (sub-) shell closure (the 
subscript $b$ denotes that these are boson Hamiltonians). 
In this work, their form is
\numparts
\begin{eqnarray}\label{eq:H_b_A}
\hat H_b^{\rm A} &= \epsilon^{\rm (A)}_d \hat n_d + 
\kappa^{\rm (A)} \hat Q_\chi \cdot \hat Q_\chi~,\\
\label{eq:H_b_B}
\hat H_b^{\rm B} &= \epsilon^{\rm (B)}_d \hat n_d + 
\kappa^{\rm (B)} \hat Q_\chi \cdot \hat Q_\chi + 
\kappa^{\rm \prime (B)} \hat L\cdot \hat L + \Delta~,
\end{eqnarray}
\endnumparts
where $\Delta$ is the off-set energy between 
configurations A and B, the quadrupole operator is 
$\hat Q_\chi = d^\dag s+s^\dag \tilde d\!+\! \chi (d^\dag 
\tilde d)^{(2)}$, where $\tilde d_\mu \eq
(-)^\mu d_{-\mu}$, and the mixing term is $\hat W_{\rm b} 
\eq \omega [(d^\dag d^\dag)^{(0)} + (s^\dag)^2] + 
\rm{H.c.}$, where H.c. stands for Hermitian conjugate.

The Hamiltonian $\hat H_b^i$, with $i=\rm A,B$, can 
interpolate between three forms of dynamical symmetry (DS) 
limits using the control parameters $(\epsilon^{(i)}_d, 
\kappa^{(i)}, \chi)$ ($\kappa^{\prime(\rm B)}$ and 
$\Delta$ of $\hat H_b^{\rm B}$ do not affect the DS 
structure), where U(6) serves as its spectrum generating 
algebra and SO(3) its symmetry algebra
\begin{equation}\label{eq:ds-chains}
\rm{U(6)} \supset \left\{\begin{array}{cc}
	  &\mkern-18mu\rm{U(5)} \supset \rm{SO(5)} \supset 
	  \rm{SO(3)} \\
      &\mkern-72mu\rm{SU(3)} \supset \rm{SO(3)} \\
	  &\rm{SO(6)} \supset \rm{SO(5)} \supset 
	  \rm{SO(3)}
\end{array}\right.
\end{equation}
In such a case, 
the spectrum is completely solvable and resembles known 
paradigms of collective motion: spherical vibrator [U(5)], 
axial deformed rotor [SU(3)] and $\gamma$-soft deformed 
rotor [SO(6)].
\subsubsection{Boson $E2$ transitions operator.}
For $E2$ transitions, the boson operator reads
\begin{equation}\label{eq:Te2-b}
\hat{T}_b(E2) =
e^{(\rm A)}\hat Q^{(N)}_{\chi} + e^{(\rm B)}\hat 
Q^{(N+2)}_{\chi}~,
\end{equation}
where $e^{(\rm A)},e^{(\rm B)}$ are the 
boson effective charges for configuration A and B, 
respectively, and superscript $(N)$ denotes a projection 
onto the $[N]$ boson space.

\subsubsection{Boson wave functions.}
The eigenstates $\ket{\Psi;L}$ of the boson Hamiltonian 
with angular momentum $L$, are linear combinations of the 
wave functions, $\Psi_{\rm A}$ and $\Psi_{\rm B}$, in the 
two spaces $[N]$ and $[N+2]$, $\ket{\Psi; L} = 
a\ket{\Psi_{\rm A}; [N], L} + b\ket{\Psi_{\rm B}; 
[N\!+\!2], L}$, with $a^{2} + b^{2} = 1$.
Each part of the wave functions with a given boson number 
$N$ and angular momentum $L$ can be expanded in terms of 
the DS bases of the IBM in the following manner
\begin{eqnarray}\label{eq:wf-ds}
\ket{\Psi; [N],L} & = \sum_{n_d,\tau,n_\Delta} 
C^{(N,L)}_{n_d,\tau,n_\Delta}\ket{N,n_d,\tau,n_\Delta,L} 
 & \qquad\left[\rm{U(5)}\right]\nonumber\\
 				  & = \sum_{(\lambda,\mu),K} 
C^{(N,L)}_{(\lambda,\mu),K}\ket{N,(\lambda,\mu),K,L} 
 & \qquad\left[\rm{SU(3)}\right]\nonumber\\
				  & = \sum_{\sigma,\tau,n_\Delta} 
C^{(N,L)}_{\sigma,\tau,n_\Delta}\ket{N,\sigma,\tau, 
n_\Delta,L}  & \qquad\left[\rm{SO(6)}\right]
\end{eqnarray}
where $N,n_d,(\lambda,\mu),\sigma,\tau,L$ label the 
irreducible representations of U(6), U(5), SU(3), SO(6), 
SO(5) and SO(3), respectively, and $n_\Delta,K$ are 
multiplicity labels. The coefficients $C^{(N,L)}_\alpha$, 
with quantum numbers $\alpha$, give the weight of each 
component in the wave function. They can therefore give us 
the probability of having definite quantum numbers of a 
given symmetry. For example, for the U(6) DS the boson 
number $N$, which reveals the amount of normal-intruder 
mixing, and for the U(5) DS the $d$-boson number, $n_d$, 
which reveals the amount of deformation
\numparts
\begin{eqnarray}
\label{eq:prob_ibm}
P^{(N_i,J)} & =~~ 
\sum_{n_d}P^{(N_i,L)}_{n_d}~;\quad i = {\rm A,B}\quad
\left(P^{(N_{\rm A},L)} + P^{(N_{\rm B},L)} = 1\right)
\\
\label{eq:prob_ibm_nd}
P^{(N_i,L)}_{n_d} & = 
\sum_{\tau,n_\Delta,L}|C^{(N_i,L)}_{n_d, \tau, 
n_\Delta,L}|^2~,
\end{eqnarray}
\endnumparts
where $N_{\rm A} = N$ and $N_{\rm B}=N+2$.
\subsection{IBFM-CM}
The IBFM for a single shell model configuration has been 
widely used \cite{IachelloVanIsackerBook} to describe 
low-lying collective states of odd-mass nuclei, where a 
fermion is coupled to the boson Hamiltonian, which serves 
as a core. For two shell model configurations, 0p-0h and 
2p-2h, the IBFM-CM framework \cite{Gavrielov2022c, 
Gavrielov2023a} Hamiltonian can be written in the following 
matrix form
\begin{eqnarray}\label{eq:ham-bf}
\fl\hat H
= \hat H_{\rm b} + \hat H_{\rm f} + \hat V_{\rm bf} 
\nonumber\\
\fl \qquad= 
\left [\begin{array}{cc}
\hat H_b^{\rm A}(\xi^{\rm(A)}) & \hat W_b(\omega) \\ 
\hat W_b(\omega)  & \hat H_b^{\rm B}(\xi^{\rm(B)})
\end{array}\right] 
+ 
\left [\begin{array}{cc}
\sum_j \epsilon_j \hat n_j &  0\\ 
	0  & \sum_j \epsilon_j \hat n_j
\end{array}\right] 
+
\left [\begin{array}{cc}
  \hat V^{\rm A}_{bf}(\zeta^{(\rm A)}) &
 \hat{W}_{bf}(\omega_j)\\
\hat{W}_{bf}(\omega_j) &
\hat V^{\rm B}_{bf}(\zeta^{(\rm B)})j
\end{array}\right]
\end{eqnarray}

The first part of $\hat H$ is the boson core with entries 
from Equations~\eref{eq:H_b_A}-\eref{eq:H_b_B}, the second 
part 
is the fermion Hamiltonian with single particle energies 
$\epsilon_j$ and the fermion number operator $\hat n_j$, 
the third part is the Bose-Fermi interaction. The latter 
has off diagonal parts that combine boson and fermion 
mixing terms and a diagonal part that typically takes the 
following form of monopole, quadrupole and exchange 
interactions
$\hat V^{(i)}_{\rm bf}(A_0,\Gamma_0,\Lambda_0) = V^{{\rm 
MON}(i)}_{\rm bf}(A_0) + V^{{\rm QUAD}(i)}_{\rm 
bf}(\Gamma_0) + V^{{\rm EXC}(i)}_{\rm bf}(\Lambda_0)$, 
with $i=\rm A,B$ \cite{IachelloVanIsackerBook}. These 
Bose-Fermi interactions depend on the strengths $A_0$, 
$\Gamma_0$ and $\Lambda_0$, respectively, given from the 
microscopic theory of the IBFM (see \cite{Gavrielov2023a} 
for more details).

\subsubsection{Bose-Fermi $E2$ transitions}
For $E2$ transitions, the boson-fermion operator reads
\begin{equation}\label{eq:TsigL}
\hat{T}(E2) =
\hat{T}_b(E2) + \hat T_f(E2) ~,
\end{equation}
where the boson part, $\hat T_b(E2)$, is given in 
\Eref{eq:Te2-b}. The fermion part, $\hat T_f(E2)$, is given 
by
\begin{equation}
\hat T_{\rm f}(E2) = 
\sum_{jj^\prime}f^{(2)}_{jj^\prime}[a^\dagger_j \times 
\tilde a_{j^\prime}]^{(2)},\label{eq:te2_f}
\end{equation}
with $f^{(2)}_{jj^\prime} = 
-\frac{e_f}{\sqrt{5}}\braket{j||Y^{(2)}_{lm}||j^\prime}$, 
where $e_f$ is the effective charge for $E2$ transitions. 

\subsubsection{Bose-Fermi wave functions}
The eigenstates of the Hamiltonian \eref{eq:ham-bf}, 
$\ket{\Psi;J}$, are linear combinations of wave functions 
\eref{eq:wf-ds}, involving bosonic basis states in the two 
spaces $\ket{[N],\alpha,L}$ and $\ket{[N+2],\alpha,L}$, 
where $\alpha$ denotes additional quantum numbers 
characterizing the boson basis used.
The boson ($L$) and fermion ($j$) angular momenta
are coupled to a total $J$ and the combined wave function 
takes the form
\begin{eqnarray}\label{eq:wf}
\fl\ket{\Psi;J} \eq
\sum_{\alpha,L,j}C^{(N,J)}_{\alpha,L,j}
\ket{\Psi_{\rm A};[N],\alpha,L,j;J} 
+ \sum_{\alpha,L_,j}C^{(N+2,J)}_{\alpha,L,j}
\ket{\Psi_{\rm B};[N+2],\alpha,L,j;J}~,
\end{eqnarray}
For such a wave function, similarly as in 
Equations~\eref{eq:prob_ibm} and \eref{eq:prob_ibm_nd}, it 
is possible to examine the probability of the quantum 
labels of the U(6) boson DS, $N$, and the U(5) boson DS, 
$n_d$,
\numparts
\begin{eqnarray}
\label{eq:prob_ibfm}
P^{(N_i,J)} & =~~ 
\sum_{n_d}P^{(N_i,J)}_{n_d}~,\quad i = {\rm A,B}\quad
\left(P^{(N_{\rm A},J)} + P^{(N_{\rm B},J)} = 1\right)
\\
\label{eq:prob_ibfm_nd}
P^{(N_i,J)}_{n_d} & = 
\sum_{\tau,n_\Delta,j,L}|C^{(N_i,J)}_{n_d, 
\tau, 
n_\Delta,L,j}|^2~
\end{eqnarray}
\endnumparts

\section{IQPTs in the zirconium and niobium chains}
\begin{figure}
\centering
\includegraphics[width=0.7\linewidth]{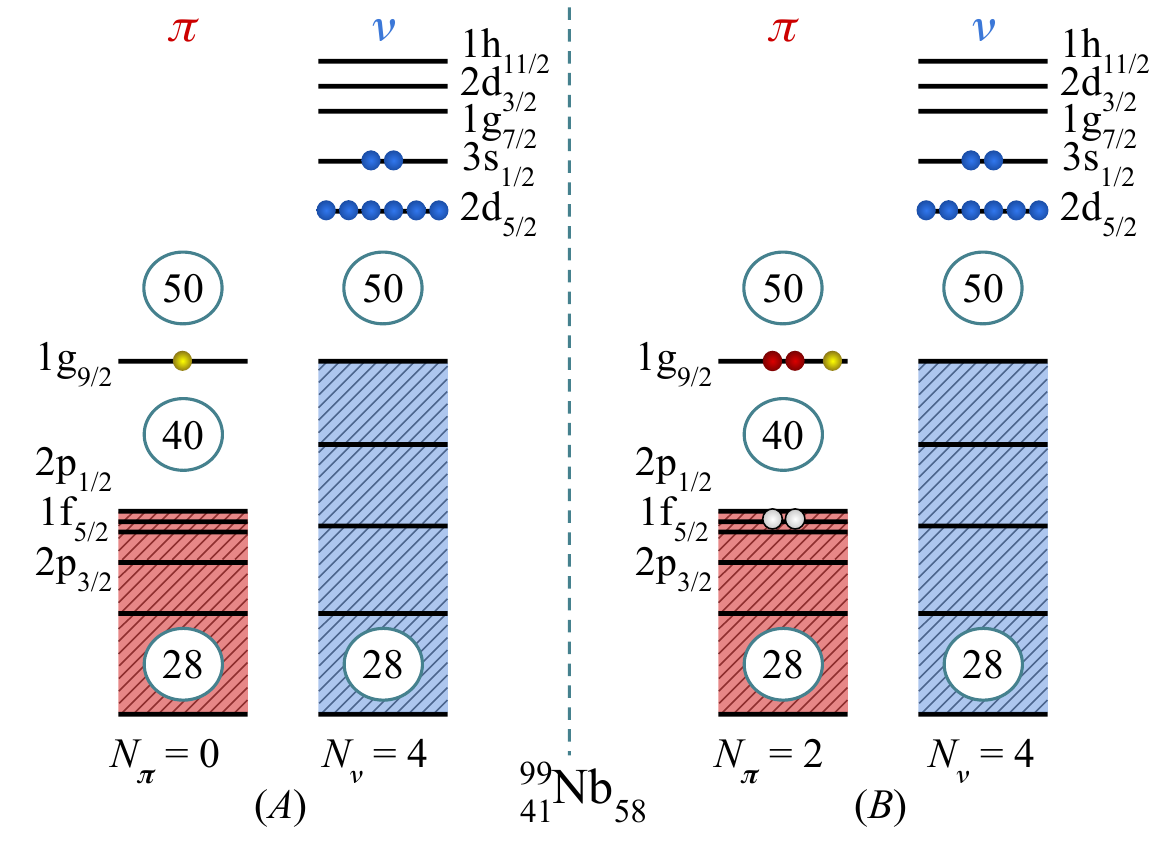}
\caption{Schematic representation of the two coexisting 
shell-model configurations (A and B) for 
$^{99}_{41}$Nb$_{58}$, where $^{98}_{40}$Zr$_{58}$ serves 
as a core. The corresponding numbers of proton bosons 
($N_\pi$) and neutron bosons ($N_\nu$), are listed for each 
configuration and $N=N_\pi + N_\nu$. \label{fig:ibfm-model}}
\end{figure}

To describe the $_{40}$Zr isotopes in the IBM-CM framework, 
we consider $_{40}^{90}$Zr as a core and valence neutrons 
in the 50--82 major shell.
The normal A configuration corresponds to having no 
active protons above $Z\!=\!40$ sub-shell gap, and the 
intruder B configuration corresponds to two-proton 
excitation from below to above this gap, creating 2p-2h 
states (see \cite{Gavrielov2022} for more details). To 
describe the $_{41}$Nb isotopes in the IBFM-CM framework we 
couple a proton to the respective $_{40}$Zr cores with 
neutron number 52--62. In the present contribution, we 
focus on the positive-parity states in the Nb isotopes, 
which reduces to a single-$j$ calculation with the 
$1g_{9/2}$ orbit (a multi-$j$ calculation for 
negative-parity states can be found in 
\cite{Gavrielov2023a}). The model space for $^{99}$Nb is 
shown schematically in \Fref{fig:ibfm-model}, where 
$^{98}$Zr serves as the boson core.

In order to understand the change in structure of the
Zr and Nb isotopes, it is insightful to examine the 
evolution of observables along the chain. In this 
contribution, the observables include energy levels, 
two-neutron separation energies and $E2$ transition rates.

\subsection{Evolution of energy levels}
\begin{figure}
\centering
\includegraphics[width=0.7\linewidth]{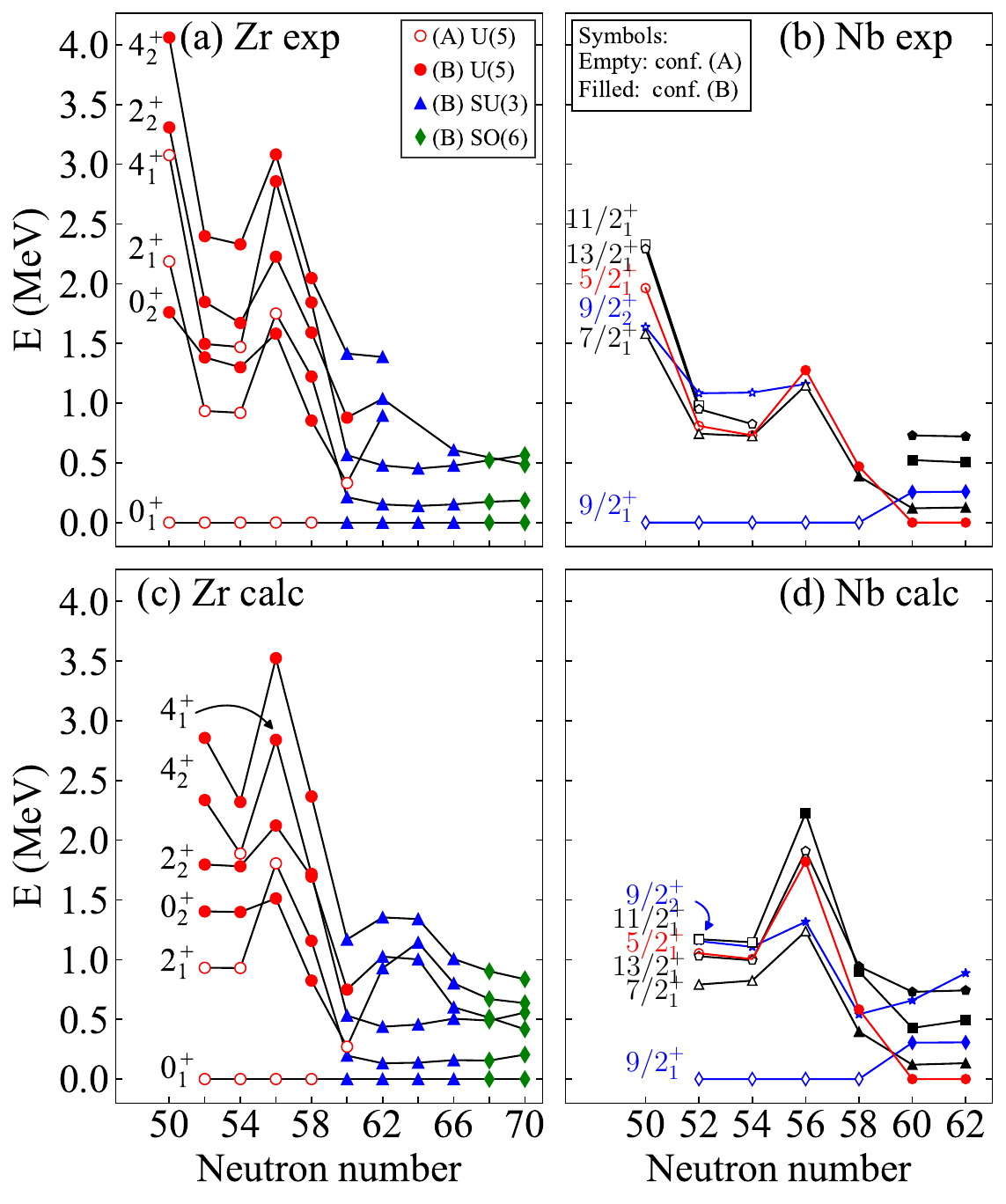}
\caption{Panels (a) and (c): comparison between 
experimental and calculated energy levels of Zr isotopes. 
Panels (b) and (d): comparison between energy levels of Nb 
isotopes.
Empty (filled) symbols indicate a states dominated by the 
normal A configuration (intruder B configuration), with 
assignments based on \Eref{eq:prob_ibm} (Zr) and 
\Eref{eq:prob_ibfm} (Nb). 
In panels (a) and (c), the shape of the symbols
[$\circ,\triangle,\diamondsuit$], indicates the 
closest dynamical symmetry [U(5), SU(3), SO(6)] to the 
level considered, based on \Eref{eq:prob_ibm_nd}.
Note that the calculated values start at neutron number 52, 
while the experimental values include the closed shell at 
50. References for the data are given at
\cite{Gavrielov2022} (for Zr) and \cite{Gavrielov2023a} 
(for Nb).\label{fig:levels}}
\end{figure}
A comparison between experimental and calculated levels of 
Zr and Nb is shown in \Fref{fig:levels}, along with 
assignments to configurations based on 
\Eref{eq:prob_ibm}, for Zr, and \Eref{eq:prob_ibfm}, for 
Nb. For Zr, we also add symbols denoting the most dominant 
DS component in the wave function based on the  
decompositions of \Eref{eq:prob_ibm_nd}, for each state 
(the definitions to the other SU(3) and SO(6) DS 
decompositions can be found in \cite{Gavrielov2022}).

For Zr isotopes, panels (a) and (c), in the region between 
neutron number 50 and 56, there appear to be two 
configurations, one spherical (seniority-like), A, and 
one weakly deformed, B, as evidenced by the ratio 
$R_{4/2}$, which is at 52--56, $R^{\rm (A)}_{4/2}\cong 1.6$ 
and  $R^{\rm (B)}_{4/2} \cong 2.3$. Their trend decreases 
at neutron number 52--54, away from the closed shell, and 
rise again at 56 due to the $\nu(2d_{5/2})$ subshell 
closure.
From neutron number 58, there is a pronounced drop in energy
for the states of configuration~B and at 60, the two 
configurations exchange their role indicating a Type~II 
QPT. At this stage, the intruder B configuration appears 
to be at the critical point of a U(5)-SU(3) Type~I QPT, as 
evidenced by the low value of the excitation energy of the 
first excited $ 0^+ $ state of this configuration
(the $0^{+}_3$ state in $^{100}$Zr shown in 
Figure 3 of \cite{Gavrielov2019}). 
Beyond neutron number 60, the intruder configuration~(B) 
is strongly deformed, as evidenced by the small value of the
excitation energy of the state $2_{1}^{+}$, 
$E_{2_{1}^{+}}\!=\!139.3$ keV and by the ratio $R^{\rm 
(B)}_{4/2}\!=\!3.24$ in $^{104}$Zr. At still larger
neutron number 66, the ground state band becomes $\gamma 
$-unstable (or triaxial) as evidenced by the close 
experimental energy of the states $2_{2}^{+}$ and 
$4_{1}^{+}$, $E_{2_{2}^{+}}\!=\!607.0$~keV, 
$E_{4_{1}^{+}}\!=\!476.5$ keV, in $^{106}$Zr, and 
especially by the experimental $ E_{4^+_1}\!=\!565$~keV and 
$ E_{2^+_2}\!=\!485$ keV in $^{110} $Zr, a signature of the 
SO(6) symmetry. In this region, the ground state 
configuration undergoes a crossover from SU(3) to SO(6).

For Nb isotopes, panels (b) and (d), we see a quintuplet of 
states ($5/2^+_1, 7/2^+_1, 11/2^+_2, 13/2^+_1, 9/2^+_2$). 
For neutron number 52--56, their energy varies due to the 
filling of the $\nu(2d_{5/2})$ shell. At neutron number 58, 
there is a pronounced drop in energy 
for the states of the B~configuration, due to the onset of 
deformation. At 60, the two configurations cross, 
indicating a Type~II QPT, and the ground state changes from 
$9/2^+_1$ to $5/2^+_1$, becoming the bandhead of a 
$K=5/2^+$ rotational band composed of $5/2^+_1, 
7/2^+_1,9/2^+_1, 11/2^+_1, 13/2^+_1, \ldots$ states. At 
this point the intruder B configuration also undergoes a 
Type~I QPT from a weak to a strong coupling 
scenario. Beyond neutron number 60, the intruder 
B~configuration remains strongly deformed and the band 
structure persists. The above trend is similar to that 
encountered in the even-even $_{40}$Zr cores.

The comparability to the even-even Zr cores comes from the 
fact that four of the states in \Fref{fig:levels},
($5/2^+_1, 7/2^+_1, 11/2^+_2, 13/2^+_1$), have the same 
trend as the $2^+_1$ state of the adjacent Zr isotopes and 
one state, ($9/2^+_2$), like the $0^+_2$. For 52--58, the 
correspondence occurs since the Zr $L=2^+_1$ state is 
coupled to single $j=9/2$ orbit through a weak coupling 
scenario giving a total angular momentum of $|L-j| \leq J 
\leq |L+j|$ (note that the $J=9/2^+$ which is part of this 
quintuplet is the $9/2^+_3$ and is not shown in the 
figure). In such a scenario, the ``center of 
gravity''~\cite{Lawson1957} of the Nb multiplet is 
centered around the energy of the $2^+_1$, as described in 
\cite{Gavrielov2023a}. The same occurs for the $9/2^+_2$ of 
Nb and the $0^+_2$ of Zr.

\subsection{Evolution of configuration and symmetry content}
\begin{figure}
\centering
\begin{overpic}[width=0.7\linewidth]{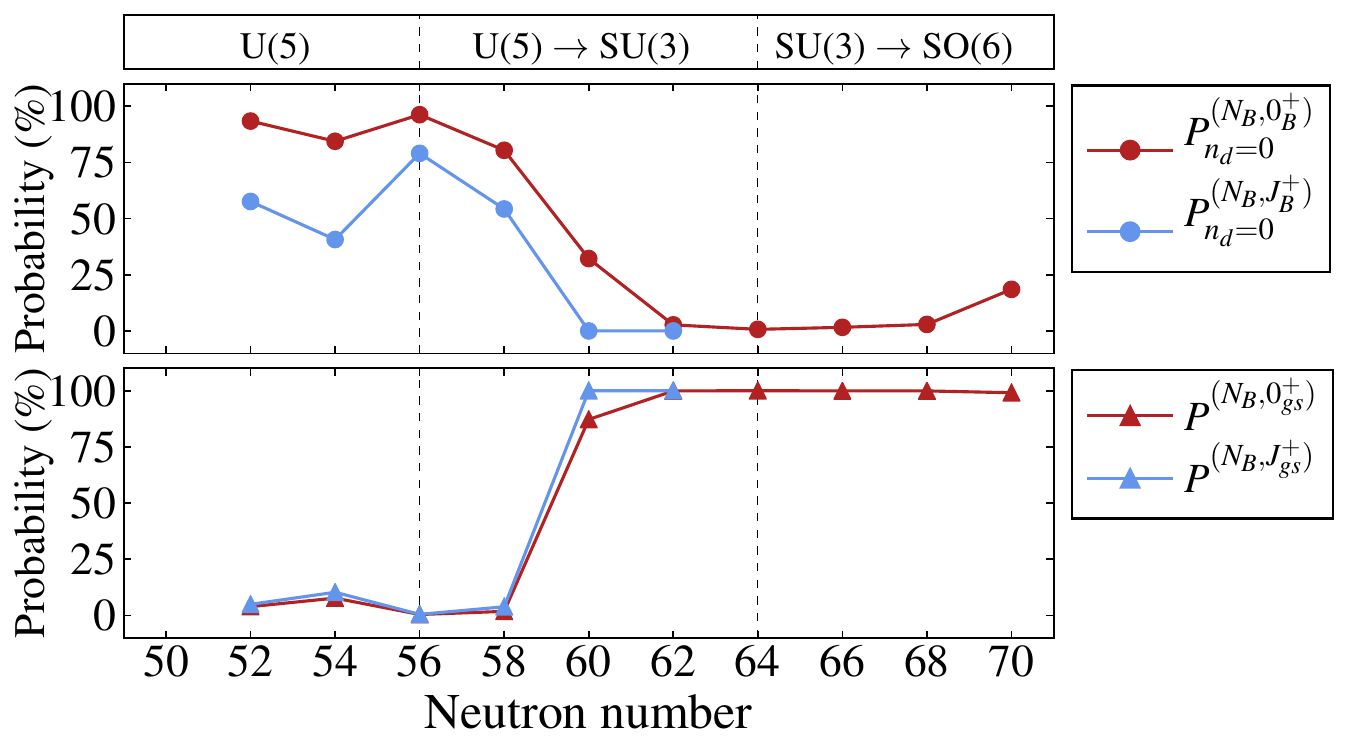}
\put (10,25) {\scriptsize\fcolorbox{black}{white}{(a) 
Intruder component}}
\put (48.7,46) {\scriptsize\fcolorbox{black}{white}{(b) 
$n_d=0$ component}}
\end{overpic}
\caption{Percentage of (a) the intruder B component, 
Equations \eref{eq:prob_ibm} and \eref{eq:prob_ibfm}, (b) 
the boson U(5) $n_d=0$ component, Equations 
\eref{eq:prob_ibm_nd} and \eref{eq:prob_ibfm_nd}, of 
$^{52-110}$Zr (red lines) and $^{93-103}$Nb (blue lines) 
isotopes, respectively. On top of panel (b) shown are the 
boson DS indicating the transitions in symmetry within the 
intruder B configuration.
For Nb isotopes, the ground state $J^+_{gs}$ ($J^+_B$) is 
$9/2^+_1$ ($9/2^+_2$) for $^{93-99}$Nb and $5/2^+_1$ 
($5/2^+_2$) for $^{101,103}$ Nb and \label{fig:decomp}}
\end{figure}
To understand the occurrence of both Type~II and I QPTs in 
the chain of Zr and Nb isotopes, we examine explicitly the 
evolution of the intruder and of the U(5) symmetry 
components of the wave function.

For the Type~II QPT, we examine in \Fref{fig:decomp}(a) the 
intruder component of the wave function of the ground state 
of Zr ($P^{(N_{\rm B},0^+_{gs})}$) and Nb ($P^{(N_{\rm 
B},J^+_{gs})}$) isotopes, Equations~\eref{eq:prob_ibm} and 
\eref{eq:prob_ibfm}, respectively. We observe the same 
trend, where from neutron number 52 to 58 this component is 
small and at 60 it jumps to about 87\% in Zr and 99\% in 
Nb and remains at about 99\% for the heavier isotopes in 
both chains. This is the occurrence of the Type~II QPT, 
where the ground state changes its configuration content 
from normal (at neutron number 58) to intruder (at 60).

For the Type~I QPT within configuration B, we examine in 
\Fref{fig:decomp}(b) the $n_d=0$ component of the 
configuration B part of $0^+_B=0^+_2$ for $^{92-98}$Zr and 
$0^+_B=0^+_1$ for $^{100-110}$Zr ($P^{(N_{\rm 
B},0^+_B)}_{n_d=0}$) and of $J^+_B=9/2^+_2$ for 
$^{93-99}$Nb and $J^+_B=5/2^+_1$ for $^{101-103}$Nb 
($P^{(N_{\rm B},J^+_B)}_{n_d=0}$). 
The circles represent the percentage of the U(5) 
$n_d\!=\!0$ component in the wave function, 
Equations~\eref{eq:prob_ibm_nd} and \eref{eq:prob_ibfm_nd}. 
For Zr, it is large ($\approx90\%$) for neutron number 
52--58 and drops drastically ($\approx30\%$) at 60. The 
drop means that other $n_d\!\not=\!0$ components are 
present in the wave function and therefore this state 
becomes deformed. Above neutron number 60, the $n_d\!=\!0$ 
component drops almost to zero (and rises again a little at 
70), indicating the state is strongly deformed.
For Nb we see the exact same trend, however, the total 
value of $P^{(N_{\rm B},J^+_B)}_{n_d=0}$ is smaller, 
indicating that for odd-mass nuclei coupling the fermion to 
the boson core increases deformation.

\subsection{Evolution of two neutron separation energies}
\begin{figure}
\includegraphics[width=1\linewidth]{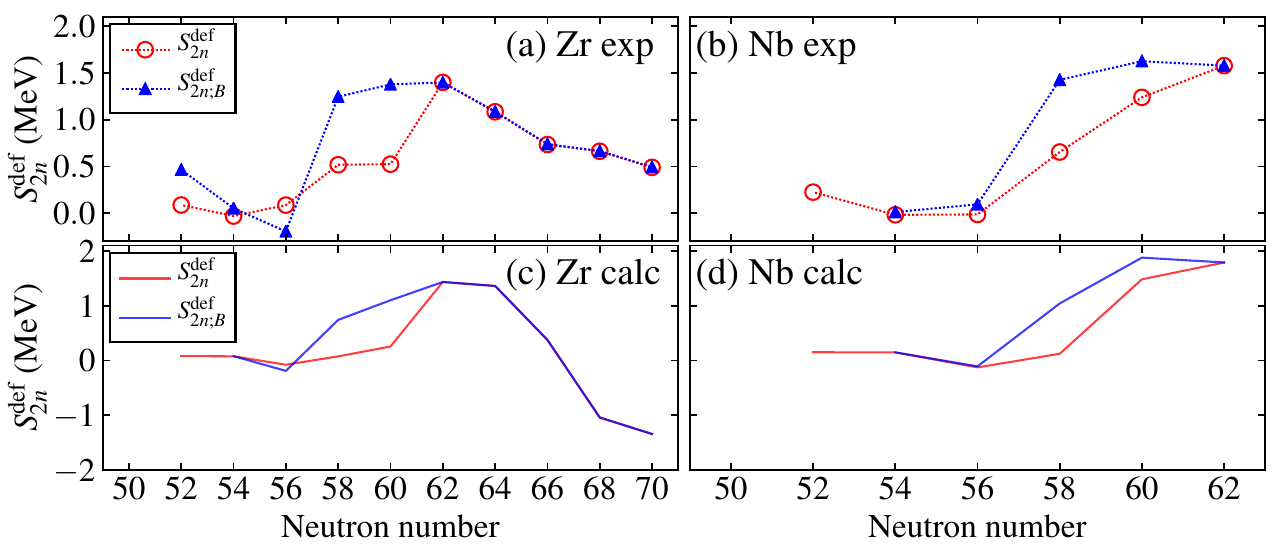}
\caption{Comparison of the deformed part of the two neutron 
separation energies, ($S^{\rm def}_{2n}$), between Zr and 
Nb experiment (a) and (b) \cite{Wang2021} and Zr and Nb 
calculation (c) and (d), respectively. 
\label{fig:s2n-def}}
\end{figure}
An observable that portrays both types of QPTs is two 
neutron separation energy, defined as 
\begin{equation}\label{eq:s2n_emp}
S_{2n} = 2M_n + M(N-2,Z) - M(N,Z),
\end{equation}
where $M(N,Z)$ is the mass of a nuclei with $N,Z$ 
neutrons and protons, respectively, and $M_n$ is the 
neutron mass.
In the IBM and IBFM, it is convenient to transcribe the 
$S_{2n}$ as
\begin{equation}\label{eq:s2n}
S_{2n} = -\tilde A - \tilde B N_v \pm S^{\rm def}_{2n} - 
\Delta_n,
\end{equation}
where $N_v$ is half the number of valence particles in the 
boson core and $S^{\rm def}_{2n}$ is the contribution of 
the deformation, obtained by the expectation value of the 
Hamiltonian in the ground state. The $+$ sign applies to 
particles and the $-$ sign to holes. The $\Delta_n$ 
parameter takes into account the neutron subshell closure 
at 56, $\Delta_n=0$ for 50--56, and $\Delta_n=2$~MeV for 
58--70 for Zr (58--62 for Nb). Its value is adapted from 
\cite{Barea2009}.
For the Zr (Nb) isotopes, the chosen values in 
\cref{eq:s2n} are $\tilde A = -17.25$ ($-16.5$) and $\tilde 
B = 0.758$~MeV. The value of $\tilde A$ is taken to fit 
$^{90}$Zr ($^{91}$Nb), and the value of $\tilde B$ is taken 
from a fit to the binding energies of $^{92,94,96}$Zr.
In \Fref{fig:s2n-def}, the Zr and Nb experimental and 
calculated deformed part, $S^{\rm def}_{2n}$ 
\cite{Petrellis2011a, Iachello2011b}, are shown in red 
open circles and lines, respectively. $S^{\rm def}_{2n}$ is 
obtained by subtracting the linear part and $\Delta_n$ from 
the experimental and calculated $S_{2n}$. One can clearly 
see the onset of deformation going from neutron number 
52--56, where $S^{\rm def}_{2n}$ is close to zero, to 
58--62, where it jumps and rises in both Zr and Nb 
isotopes. For Zr, we also see after neutron number 62 a 
gradual decrease in $S^{\rm def}_{2n}$ due to the crossover 
towards SO(6) symmetry.

In order to denote the occurrence of both 
Type~I and II QPTs, in addition to \Eref{eq:s2n}, 
it is also possible using \Eref{eq:s2n_emp} to estimate two 
neutron separation energies for excited states, by using 
the mass of an excited state $M(N,Z) \!\equiv\! 
M_{exc}(N,Z) = M_{gs}(N,Z) + E_{exc}(N,Z)$, where 
$M_{gs}(N,Z)$ is the mass for the ground state and 
$E_{exc}(N,Z)$ is the energy 
of the excited state \cite{Gavrielov2023a}. We choose the 
lowest configuration~B state for this case in both Zr 
and Nb isotopes. The experimental and calculated results, 
$S^{\rm def}_{2n;B}$, are given in blue triangles and 
lines, in \Fref{fig:s2n-def}. For both Zr and Nb isotopes, 
it is seen that for neutron number 54--56 $S^{\rm 
def}_{2n;B}$ is close to zero, then at 58 it jumps due to 
the onset of deformation, then it flattens at 60. 
This behavior denotes the \mbox{Type~I}~QPT of shape 
evolution within configuration~B of Zr (Nb) from spherical 
(weak coupling) to axially-deformed (strong coupling).
For neutron number 54--56, $S^{\rm def}_{2n;B}$ 
(triangles) is close to the value of $S^{\rm def}_{2n}$ 
(circles), as configuration~B is more spherical. At 58, 
there is a larger jump than $S^{\rm def}_{2n}$ since 
configuration~B is more deformed than A, which 
continues at 60.
For 62 (and 64--70 for Zr), both $S^{\rm def}_{2n;B}$ and 
$S^{\rm def}_{2n}$ coincide since the ground state is 
configuration~B, which denotes the \mbox{Type~II} QPT.

\subsection{Evolution of $E2$ transition rates}
\definecolor{GreenNoam}{rgb}{0,0.5,0}
\begin{figure}[t]
\centering
\includegraphics[width=1\linewidth]{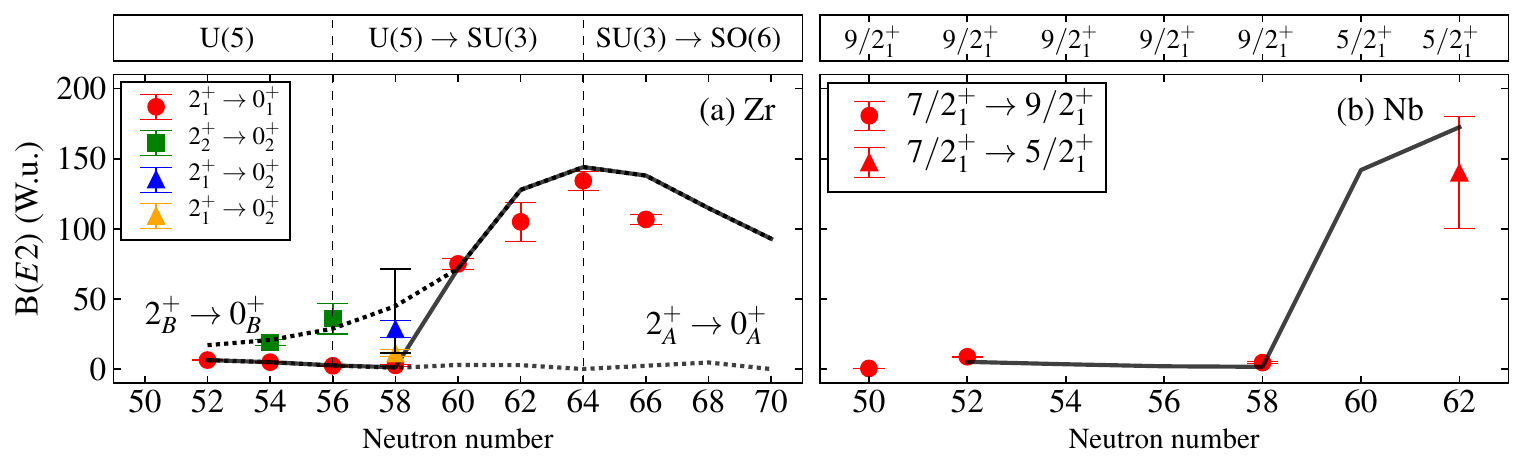}
\caption{$B(E2)$ values in W.u. (a) for $2^+ \to 0^+$ 
transitions in the Zr chain and (b) for $7/2^+_1 \to 
J^+_{gs}$ in the Nb chain, where $J^+_{gs}$ is given in the 
panel above (b). On top of panel (a) shown are the 
boson dynamical symmetries indicating the transitions in 
symmetry within the intruder B configuration.
The solid line (symbols 
${\color{red}\CIRCLE}, ~{\color{GreenNoam}\blacksquare}, 
~{\color{blue}\blacktriangle}, 
~{\color{orange}\blacklozenge}$) denote calculated results 
(experimental results). In (a), dotted lines denote 
calculated $E2$ transitions within a configuration. 
References for the data can be found in 
\cite{Gavrielov2022}.\label{fig:be2}}
\end{figure}
The occurrence of both Type~I and II QPTs are also stressed 
by an analysis of $B(E2)$ values.
As shown in \cref{fig:be2}(a) for Zr, the calculated 
$2^+_A\to 0^+_A$ transition rates coincide with the 
empirical $2^+_1\to 0^+_1$ rates for neutron numbers 
52--56. The calculated $2^+_B\to 0^+_B$ transition rates 
coincide with the empirical $2^+_2\to 0^+_2$ rates for 
neutron numbers 52--56, with the empirical $2^+_1\to 0^+_2$ 
rates at 58 and with the empirical $2^+_1\to 
0^+_1$ rates at 60--64. The large jump in 
$\be{2}{1}{0}{1}$ between 58 and 60 reflects 
the passing through a critical point, common to a Type II 
QPT involving a crossing of two configurations.
The further increase in $\be{2}{1}{0}{1}$ for neutron 
numbers 60--64 is as expected from a U(5)-SU(3) Type~I QPT 
and reflects an increase in the deformation in a spherical 
to deformed shape-phase transition within configuration~B. 
The subsequent decrease from the peak at neutron number 64 
towards 70 is expected from an SU(3) to SO(6) crossover.
The same trend at neutron number 52--62 in the $B(E2)$ 
value Zr can be observed in \cref{fig:be2}(b) for the Nb 
isotopes. Although there is less data, the $B(E2)$ values 
from the first excited to the ground state are small for 
neutron number 52--58 and at 60 there is large jump, which 
continues to increase towards 62, possibly suggesting even 
stronger deformation than in the adjacent Zr, although the 
experimental error bar is large.

\section{Conclusions and outlook}
The general framework of the interacting boson model and 
interacting boson-fermion model with configuration mixing,
IBM-CM and IBFM-CM, respectively, has been presented, 
allowing a quantitative description of shape-coexistence 
(configuration-mixing) and related QPTs in even-even and 
odd-mass nuclei.
By employing such a comparison, odd-mass nuclei can serve 
as a further test case for shape coexistence in even-even 
nuclei by verifying whether the even-even core is the right 
one to describe the adjacent odd-mass nucleus.

A quantal analysis for the chain of the even-even $_{40}$Zr 
and odd-mass $_{41}$Nb isotopes involving positive-parity 
states was performed for neutron number 52--70 (Zr) and 
52--62 (Nb). It examined the evolution of energy levels, 
two-neutron separation energies and $E2$ transition 
rates along both chains. Special attention has been devoted 
to changes in the configuration-content and U(5) content of 
wave functions.

The results of the analysis suggest a complex phase 
structure yet similar in both chains of these isotopes, 
involving two configurations.
The two configurations cross near neutron number 60, and 
the ground state changes from configuration A to 
configuration B, demonstrating a Type~II QPT. They are 
weakly mixed and retain their purity before and after the 
crossing. Alongside the Type~II QPT, the intruder 
B configuration undergoes a spherical-U(5) (weak 
coupling) to axially-deformed-SU(3) (strong coupling) QPT 
within the boson core (Bose-Fermi Hamiltonian), with a 
critical point near $A\approx100$, thus demonstrating, 
IQPTs in both even-even and odd-mass nuclei.

\ack
The author acknowledges support from the European Union's 
Horizon 2020 research and innovation program under the 
Marie Sk\l{}odowska-Curie grant agreement No. 101107805. 
The results reported are based on work done in 
collaboration with F.~Iachello (Yale University) and 
A.~Leviatan (The Hebrew University).

\section*{References}
\bibliography{refs.bib}
\end{document}